\documentclass[]{emulateapj}
\usepackage{graphicx}
\usepackage{natbib}

\newcommand{\be}{\begin{eqnarray}}
\newcommand{\ee}{\end{eqnarray}}
\def\HeI{\ion{He}{1}}
\def\HeII{\ion{He}{2}}
\def\HeIII{\ion{He}{3}}

\begin{document}
\title{Detailed and simplified non-equilibrium helium ionization in the solar atmosphere}
\shorttitle{Non-equilibrium helium ionization}
\shortauthors{T. P. Golding, M. Carlsson and J. Leenaarts}

\author{Thomas Peter Golding}
 \affil{Institute of Theoretical Astrophysics, University of Oslo,
   P.O. Box 1029 Blindern, NO-0315 Oslo, Norway}
 \email{thomas.golding@astro.uio.no}
\author{Mats Carlsson}
  \affil{Institute of Theoretical Astrophysics, University of Oslo,
    P.O. Box 1029 Blindern, NO-0315 Oslo, Norway}
  \email{mats.carlsson@astro.uio.no}
 \author{Jorrit Leenaarts\altaffilmark{1}}
  \affil{Institute of Theoretical Astrophysics, University of Oslo,
    P.O. Box 1029 Blindern, NO-0315 Oslo,  Norway}
  \email{jorritl@astro.uio.no}  
\altaffiltext{1}{now at Institutet f\"or solfysik - Stockholms Universitet}

\begin{abstract}
Helium ionization plays an important role in the
energy balance of the upper chromosphere and transition region. 
Helium spectral lines are also often used as diagnostics of these regions.
We carry out 1D radiation-hydrodynamics simulations of the solar
atmosphere and find that the helium ionization
is mostly set by photoionization and direct collisional ionization, counteracted
by radiative recombination cascades. 
By introducing an additional recombination rate mimicking the
recombination cascades, we construct a simplified 3 level helium
model atom consisting of only the ground states. This model atom is
suitable for modeling non-equilibrium helium ionization in 3D
numerical models.
We perform a brief investigation of the formation of the \HeI\ 10830 and
\HeII\ 304 spectral lines.  Both lines show non-equilibrium features that are not 
recovered with statistical equilibrium models, and caution should
therefore be exercised when such models are used as a basis in the
interpretation of observations. 
\end{abstract}


\section{Introduction}

The gas in the solar atmosphere goes from mostly neutral in the
photosphere to highly ionized in the corona.
In the dynamic interface of these two regimes, the
chromosphere and transition region, atoms will ionize
and recombine on various timescales. The ionization will be out of
equilibrium if the ionization/recombination timescale is longer than
the typical hydrodynamic timescale. The ionization balance of
sufficiently abundant atomic species affect the energetics of the
atmosphere, as an ionized atom stores energy that would otherwise
increase the temperature of the gas.
Stellar atmosphere simulations that use the simplifying assumption of
statistical equilibrium (SE) might therefore miss potential
effects of non-equilibrium ionization.

In a numerical model of the solar atmosphere, treating ionization-recombination 
processes in detail requires solving the complete radiative transfer
problem - a non-linear, non-local problem which in 3D is too
computationally demanding for present day computers to handle. In simpler geometry, 
however, the
situation is different: \cite{carlsson_stein1992, carlsson_stein1995,
  carlsson_stein1997} carried out 1D simulations
of a dynamic solar atmosphere where the non-equilibrium ionization and
recombination of abundant elements was included, and they found that the
effects of this are indeed important for the thermodynamic structures of
the atmosphere. In a follow-up study, it was shown that the
relaxation timescales of hydrogen ionization and recombination are long compared to the dynamic 
timescales, especially in the cool post-shock phase \citep[hereafter referred to as
CS2002]{carlsson_stein2002}. 
This leads to a lower ionization degree in 
chromospheric shocks and a higher ionization degree between the shocks,
compared to the statistical equilibrium solution, since the hydrogen
populations do not have time to adjust to the rapidly changing conditions.

Based on a simplified method for treating the radiative transition rates of hydrogen
\citep{sollum1999}, an experiment with hydrogen ionization 
was carried out by \cite{leenaarts2006} in 3D. This study confirmed that 
hydrogen is out of equilibrium in the chromosphere.
The experiment was repeated by
\cite{leenaarts2007} in 2D, but this time the ionization was included also in the
equation-of-state (EOS), resulting in larger temperature variations in and between 
the shocks propagating in the chromosphere than what was found with an
EOS assuming local thermodynamic equilibrium (LTE). Non-equilibrium formation  of $\mathrm{H}_2$ was later
included in this method \citep{leenaarts2011} and is currently a part
of the Bifrost stellar atmosphere code
\citep{gudiksen2011}. 

\cite{leenaarts2011} pointed out that their
model had a temperature plateau ($\sim 10$~kK ) in the upper
chromosphere and that it most likely was associated with the LTE
treatment of helium in their EOS. Our goal is to realistically treat
the non-equilibrium ionization of both hydrogen and helium in
Bifrost. With such a model we plan to perform studies of  the
formation of the spectral lines \HeI\ 10830 (formed in the chromosphere) and \HeII\
304 (formed in the transition region), both of which are  often used diagnostics,
e.g. SDO/AIA  \citep{lemen2012}, STEREO's SECCHI/EUVI
\citep{howard2008}, VTT/TIP II: \citep{collados2007}, NST/NIRIS
\citep{cao2012}.  

The 10830 line is an absorption line that forms when continuum photons from the 
photosphere are scattered or absorbed in the chromosphere by neutral 
helium atoms occupying the metastable $2s \,^{3}\!S$ state. This state is 
mostly populated by recombination cascades that follow from the photoionization of
neutral helium atoms by the coronal EUV incident
radiation \citep{avrett1994, mauas2005, centeno2008}.
For this reason the line
maps out the boundaries of coronal
holes, which are regions where the coronal EUV emission is weaker
\citep{sheeley1980, harvey2002}. \HeI\ 10830 is a valuable line also
for studying active regions where the EUV emission is strong. An
example is \cite{ji2012} who used high resolution imaging data from
NST to study heating events in small scale magnetic loops. 
Another application is the study of magnetic fields:
\cite{xu2012} used inversions of the the full Stokes vectors of the
photospheric \ion{Si}{1} 10827 and the chromospheric \HeI\ 10830 to
study the magnetic field associated with an active region
filament. Owing to its diagnostic potential, the 10830 line is one of
the candidate lines for the planned Solar-C mission.

\HeII\ 304 is an optically thick line that forms in the transition
region. It is an important source of the impinging EUV radiation
absorbed by the chromosphere, hence it might be important for the
spectrum of \HeI\ \citep{andretta2003} as well as for the energy balance
of the chromosphere and transition region. 
% history of line
Its formation is still debated. As noted by \cite{jordan1975},
the helium line intensities are larger than what is predicted by
models constructed from observations of other EUV lines. This author introduced the idea of
high energy electrons mixing with cold ions, typically associated
with a steep temperature gradient, to enhance the predicted
intensity. \cite{laming1992} elaborated on a similar idea involving
high energy electrons
in a burst model to match observed intensities. An alternative
view of the 304 formation is the photoioniziation-recombination picture,
where coronal EUV incident radiation photoionizes \HeII\ leading to
a recombination cascade ultimately ending up in a 304 photon
\citep{zirin1988}. Both of these processes may be of importance.
\cite{andretta2003} showed that at least for the
quiet sun, there are not enough coronal EUV photons produced
to account for all of the 304 emission.
The line is often used for 
the study of filaments and prominences \citep{liewer2009, bi2012,
  labrosse2012}. Another recent example of its use is for the study of
spicules \citep{murawski2011} and heating events associated with them
\citep{depontieu2011}.  

In this paper we describe a series of 1D radiation-hydrodynamics
simulations similar to those of CS2002. Based on the results we derive
a simplified helium model atom suitable for treating non-equilibrium
helium ionization in 3D numerical models. Additionally, we perform an
initial investigation of non-equilibrium ioniziation effects on the 10830 and
304 spectral lines. In Section \ref{section:method} we describe the
code, simulation setup and the assumptions for the construction of the
simplified model atom, in Section \ref{sec:results} we describe our
results. Finally, in Section \ref{section:conclusions} we draw
conclusions.

\section{Method}\label{section:method}
We use the 1D radiation-hydrodynamics code RADYN 
\citep[][CS2002]{carlsson_stein1992, carlsson_stein1995, carlsson_stein1997},
which solves the equations of mass, momentum, charge 
and energy conservation, as well as the rate equations, on an adaptive grid. 
The simulations include a detailed treatment of
non-equilibrium excitation, ionization and radiative transfer from the
atomic species H, He and Ca. Other elements are also included, but
their contribution to the ionization energy and background opacity is
based on the LTE assumption and read from
a table produced by the Uppsala opacity program
\citep{gustafssen1973}. We devote special attention to the internal
energy balance equation:
\be
 e=\frac{3kT}{2} \left( n_\mathrm{e}+\sum_{i,j} n_{i,j} \right)
      +\sum_{i,j} n_{i,j} \chi_{i,j}
        \label{eq:energy}
\ee
where $e$, $n_\mathrm{e}$, $n_{i,j}$ and $\chi_{i,j}$ are the internal energy, electron
density, population density and excitation/ionization energy
corresponding to the $i$-th state of the $j$-th ion. The two terms on the right side
represent the contributions from thermal energy and
ionization/excitation energy.

Hydrogen and singly ionized calcium are
modeled with six level atoms and helium is modeled with a 33 level atom.
Each line is described with 31-101 frequency points, whereas 6-90
frequency points are used for the various continua. 
The simulations carried out in this study are very similar to the one
carried out in CS2002 with the major difference being the very detailed helium model atom included here. For more technical information about the
code, we refer to this paper and the references therein. 

\subsection{Model atoms}
The hydrogen and calcium model atoms are identical to the ones used in
CS2002.
The helium model atom is a reduced version of an atomic model
extracted from HAOS-DIPER\footnote{The HAO  Spectral Diagnostic
  Package for Emitted Radiation:
  http://www.hao.ucar.edu/modeling/haos-diper}.  
This original model has 75
energy levels: the ground state of neutral helium plus 48 excited states (14
singlet and 34 triplet) up to $n=5$, the ground state of
\HeII\ and 24 excited states up to $n=5$, and \HeIII. The energies of
the \HeI\  and \HeII\ states
are from the National Institute of Standards and Technology
(NIST)\footnote{www.nist.gov} database and \cite{sugar1979}, respectively.  
The model atom has 311
transitions: 255 lines and 56 continua. Line oscillator strengths for
 neutral helium are from NIST, and for the transitions in
\HeII\ from \cite{parpia1982}. All
photoionization cross sections are from the OPACITY
project\footnote{http://cdsweb.u-strasbg.fr/topbase/TheOP.html}. The collisional
rates of bound-bound neutral helium transitions are from
\cite{sawey1993} and the neutral bound-free rates from excited states are
modeled with Seaton's
semi empirical formula for neutrals \citep[Section 18]{allen1973}. 
For \HeII\ the bound-bound collisional rates are taken from
CHIANTI \citep{dere2009} and the bound-free collisional rates from excited
states are from  \cite{burgess1983}. The collisional ionization rate from the
ground states of both \HeI\ and \HeII\ is
from \cite{ar1985}. 

We reduce the number of levels in the atom from 75 to 33 by merging
the neutral helium singlet $n=4$ states into one representative
state. Similarly, the neutral helium singlet $n=5$ states, triplet $n=4$ states and
triplet $n=5$ states are merged into three representative states. Also the
\HeII\ $n=4$ and $n=5$ states are merged into two
representative states.
The merging of levels is done by the method described in
\cite{bard2008}, i. e. energies of merged levels are weighted
averages and transition probabilities are computed under the
assumption that the original levels have similar energies
and identical departure coefficients (ratio between population density in non-LTE and LTE).

\subsection{Boundary conditions}
\begin{figure}
  \includegraphics[width=\columnwidth]{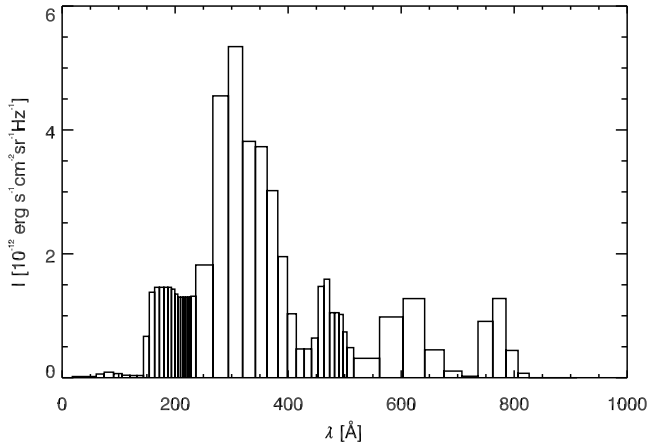}
  \caption{EUV incident radiation field at the upper boundary
  of the computational domain. The data has been binned at the same
  resolution as the employed frequency grid, while preserving the
  frequency-integrated energy flux.}
  \label{fig:fig_incrad}
\end{figure}
 The boundary conditions are the same as those in CS2002. Both the
upper and lower boundaries are transmitting. The lower boundary is
located at a fixed geometrical depth, corresponding to 480~km below
$\tau_{500}=1$ in the initial atmosphere. The incoming
characteristics at the lower boundary are prescribed, and they
result in waves that propagate up through the atmosphere.  The 
upper boundary is located at 10 Mm and has a fixed temperature of
$10^6$~K representing a corona. 
The upper boundary is irradiated by a fixed EUV incident
radiation field (see Figure \ref{fig:fig_incrad}) that represents
illumination by the corona. This radiation field is equal to that
derived in \cite{wahlstrom1994} based on data from
\cite{tobiska1991}.

\subsection{Simulations}
\label{sec:simulations}
We carried out three simulations, each running for 3600~s of solar time. The
simulations differ only in the treatment of helium. We used the following
setups: the 33-level model atom with 
non-equilibrium population densities (referred to as the NE-run),
the 33-level atom with statistical equilibrium population densities (referred to as
the SE-run) and a 3-level atom (derived in Section \ref{section:simplified_atom}) 
with non-equilibrium population densities (referred to as the NE3-run).

Several smaller simulations are carried out in addition to the three
main runs in order to determine the relaxation timescales of the helium
ionization/recombination processes. The initial atmospheres of these 
runs are snapshots very similar to those found in the NE-run (as they are 
from a simulation where a 9 level He model atom was used instead of 
the 33 level model atom). These initial atmospheres contain the
statistical equilibrium (SE) solution 
for the thermodynamic state of the atmosphere. The
temperature is then increased by 1\% and the populations are allowed to
adjust, during which all other quantities are forced to remain
constant with zero velocity.
In addition, we computed the SE solution of the perturbed atmospheres.

\begin{figure}
  \includegraphics[width=\columnwidth]{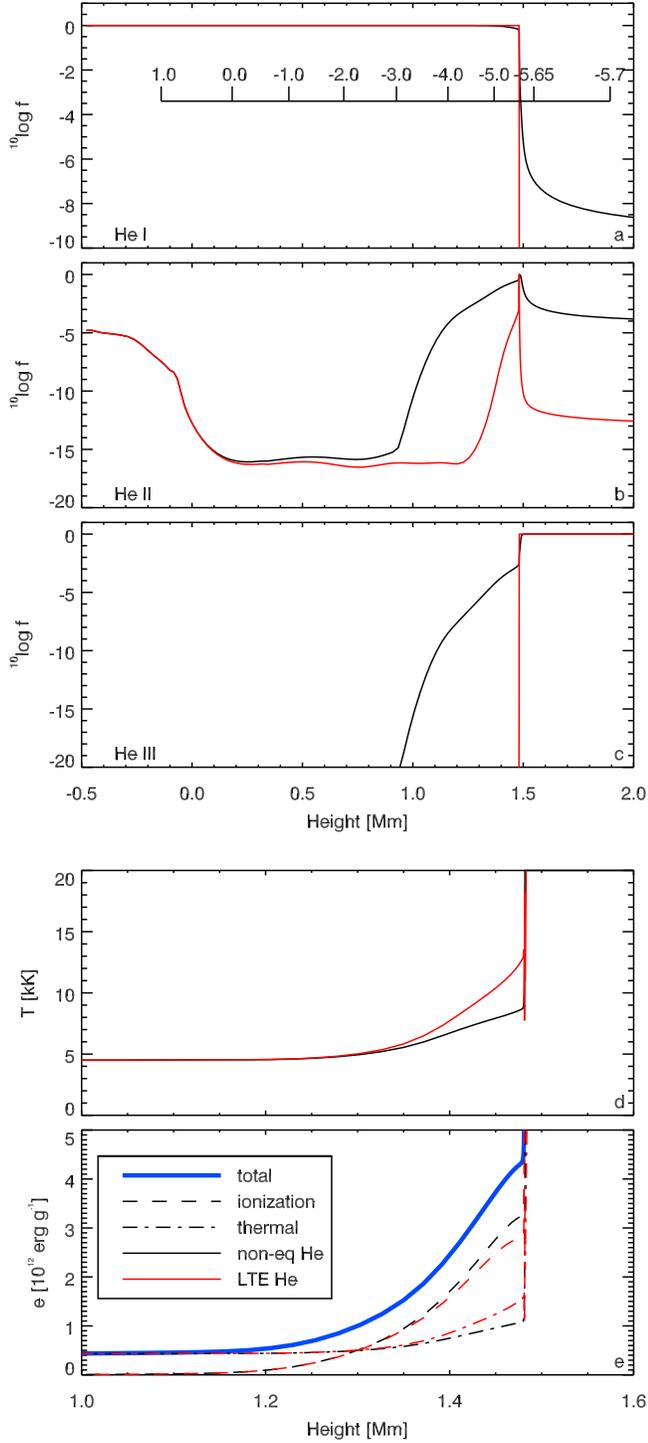}
  \caption{Panel {\it a}--{\it c}:  Helium SE (black) and LTE (red) ion fractions of the
    initial atmosphere. The $^{10}$log of the
    column mass [g cm$^{-2}$] is indicated in the upper panel.
    Panel {\it d}: comparison of the
    temperatures assuming
    SE (black) and LTE (red) helium ionization in the energy equation
    (Eq. \ref{eq:energy}). Panel {\it e}: The terms of
    Eq. \ref{eq:energy}  assuming SE (black) and LTE (red). The temperature is
    sensitive to the levels of helium ionization.}
 \label{fig:fig2_ionfracs}
\end{figure}
\begin{figure}
  \includegraphics[width=\columnwidth]{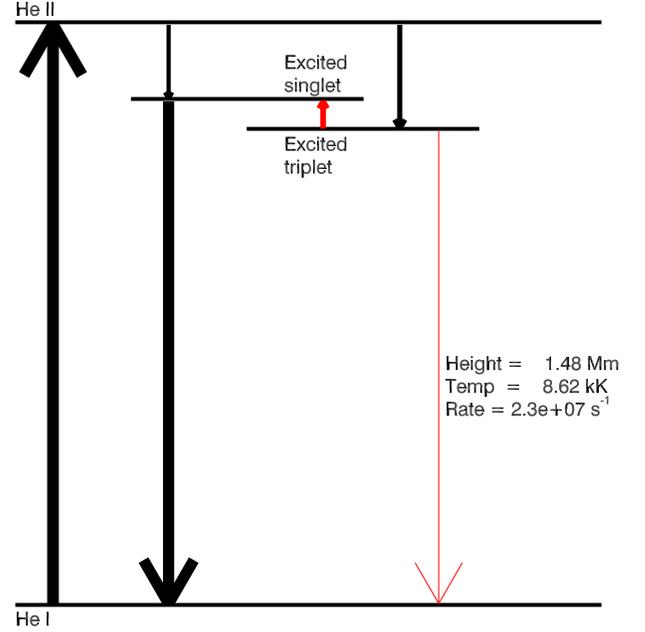}
  \caption{Net transition rates in the \HeI\ - \HeII\
    system of the initial atmosphere. The red and black arrows represent
    collisional and radiative transitions, respectively. Their
    thickness is proportional to their absolute value; the
    maximum net rate is denoted in the figure as Rate. The driver of
    the system is the photoionization from the ground state of neutral
    helium, which is balanced mainly by radiative
    recombination cascades. This picture is qualitatively  valid from the temperature
    minimum at a height of 0.9~Mm to the transition region at 1.5~Mm.}
  \label{fig:hei_transitions}
\end{figure}
\begin{figure}
  \includegraphics[width=\columnwidth]{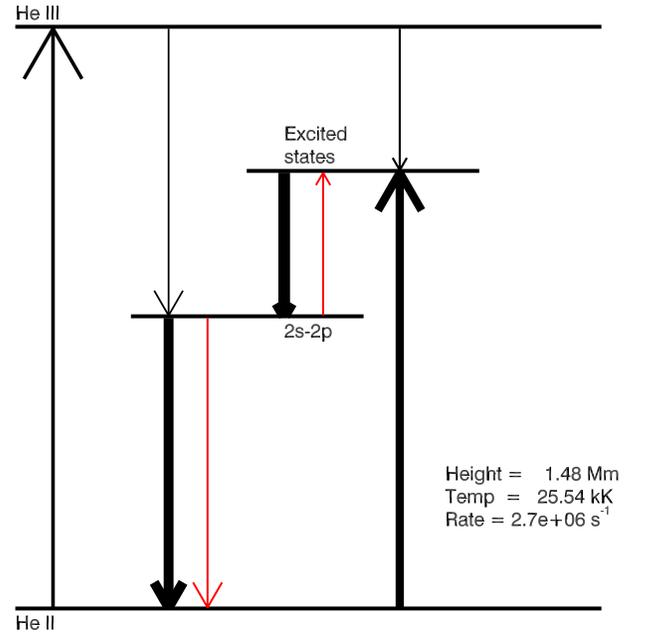}
  \caption{Net transition rates in the \HeII\ - \HeIII\ system at the
    base of the transition region in the same 
    format as Figure~\ref{fig:hei_transitions}.  Similar to
      the \HeI\ - \HeII\ system,
      photoionization from the ground state, balanced by radiative
      recombination cascades dominate in setting the ionization degree. At
      higher temperatures, collisional ionization becomes more
      important.}
  \label{fig:heii_transitions}
\end{figure}

\subsection{Initial atmosphere}\label{sec:initial_atmosphere}
The initial atmosphere for all runs is at rest and all population densities are in
SE. Figure \ref{fig:fig2_ionfracs} gives a comparison of the ion
fractions, $f=n_{\mathrm{ion}}/n_{\mathrm{total}}$, and the
corresponding LTE values.
At low heights helium is mostly neutral, and only from the
upper chromosphere and upwards ($z>1.2$~Mm) do we find \HeII\
fractions above $10^{-3}$. The \HeII\ fraction peaks somewhere in the transition region 
above which effectively all helium is in the form of \HeIII. LTE is a
decent approximation in the photosphere, but it fails to reproduce
realistic ion fractions from the chromosphere and up since LTE does
not take into account radiative transitions, whose most important
contribution is the photoionization caused by coronal EUV radiation
\citep{zirin1975}. 

Since we suspect that the ionization state of helium has an effect on the
energy balance of the atmospheric gas, we carry out a rough
numeric test of the temperature's sensitivity to the ionization state
of helium. This is done in the following way: we fix the internal energy of the initial
atmosphere, set the helium population densities to their LTE value and
then re-solve the internal energy equation (Eq. \ref{eq:energy}) together with the
charge conservation equation (as ionized helium is a significant source of
electrons). 

We compare the energy balance between thermal energy
and ionization energy in the SE and LTE cases in
the two lower panels of Figure \ref{fig:fig2_ionfracs} (note the different height
scale). Moving upwards from the photosphere, the temperature in the SE
and LTE cases starts to deviate in the upper chromosphere. 
The density of \HeII\ ions is several orders of magnitude higher in SE
than compared to what is predicted by LTE. 
This leads to a higher ionization energy, which, since the total
energy is fixed, must result in a decrease of the thermal energy.
Continuing into the transition region the
LTE temperature drops abruptly before it rises to the coronal
value. The sudden drop happens because helium in LTE goes from
neutral to fully ionized in a very short temperature interval centered
at about 20~kK. This leads to an increase in the ionization energy and a
subsequent fall in the thermal energy that is not present with helium
in SE.
This illustrates that the temperature in the solar chromosphere and
transition region is sensitive to the helium
ionization balance.

The ionization state in the upper chromosphere of the initial model atmosphere
is mostly set by
photoionization from the ground states of \HeI\ and \HeII\ followed by
a radiative recombination cascade through the 
excited states back to the ground states.  This is illustrated in Figure
\ref{fig:hei_transitions} and \ref{fig:heii_transitions}.

\subsection{Simplified model atom}\label{section:simplified_atom}
Figures~\ref{fig:hei_transitions} and \ref{fig:heii_transitions}
do not show how important the photoionization from the excited states
is, only that it is less frequent than radiative recombination. 
We make the simplifying assumption that the excited states 
act solely as intermediate steps in recombination cascades, and that
they do not serve as states from which photoionization is taking
place. Based on this assumption, we set up a model atom with only
three levels: ground-state \HeI\, ground-state \HeII\ and \HeIII. 

In addition to the ordinary collisional and radiative rates between
these levels, we add an extra recombination rate for each of the two
ions, \HeII\ and \HeIII. This extra rate is meant to model the net
recombination to excited states. 

The number of photoionizations ($n_iR_{ic}$), spontaneous 
recombinations ($n_cR_{ci}^{\mathrm{sp}}$) and induced
recombinations ($n_cR_{ci}^{\mathrm{in}}$) between
a level $i$ and its overlying continuum level $c$ 
can be expressed as
\be
  n_iR_{ic}  &=& n_i f_{ic}(J_{\nu}) \label{eq:photoionization}  \\
  n_cR_{ci}^{\mathrm{in}} &=& n_c n_\mathrm{e}
               f_{ci}^{\mathrm{in}}(T,J_{\nu})  \label{eq:indrec} \\
  n_cR_{ci}^{\mathrm{sp}} &=& n_c n_\mathrm{e}
               f_{ci}^{\mathrm{sp}}(T),  \label{eq:spontrec} 
\ee
where $n_i$ and  $n_c$  are the population densities of the
$i$-th and $c$-th level and  $J_{\nu}$ is the 
 (frequency-dependent) mean intensity. The functions $f_{ic}(J_{\nu})$, 
$f_{ci}^{\mathrm{in}}(T,J_{\nu})$ and $f_{ci}^{\mathrm{sp}}(T)$ depend
on temperature and mean intensity   
\citep{mihalas1978}. The net recombination from the 
$c$-th ionization stage to excited states of the $(c-1)$-th ionization stage, 
$P_c$,  can now be expressed as
\be 
P_c=n_c n_\mathrm{e}\sum_i(f_{ci}^{\mathrm{sp}} + f_{ci}^{\mathrm{in}})
- \sum_in_i f_{ic},
\ee
where $i$ in the sum represents the excited states. 

\begin{figure}
  \begin{center}
   \includegraphics[width=\columnwidth]{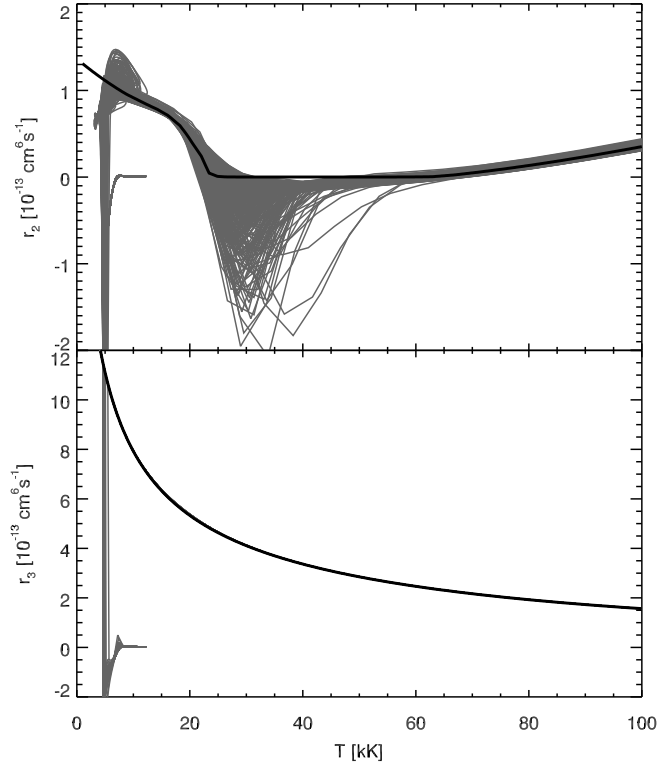}
  \end{center}
 \caption{Effective recombination rate coefficient, $r_c$, as a function
    of temperature for neutral (upper panel) and singly ionized helium
   (lower panel). The grey lines correspond to different snapshots
   from the NE-run simulation and the black line represents the value chosen for
   the model atom. Negative values
   correspond to net photoionization from excited states.}
  \label{fig:recrate_falc}
\end{figure}

We want to model this with an effective recombination rate 
$P_c=n_c n_\mathrm{e}r_c$, where $r_c$
is dependent on
temperature only. This leads to the following expression for $r_c$:
\be
   r_c = \sum_{i}(f_{ci}^{\mathrm{sp}}(T) + f_{ci}^{\mathrm{in}}(T, J_{\nu})) - \frac{\sum_in_if_{ic}(J_{\nu})}{n_c n_\mathrm{e}}.
\label{eq:effrec}
\ee
If photoionization from excited states and induced recombination are
negligible, $r_c$ will be dominated by spontaneous recombination that is dependent on temperature only. 

In  Figure \ref{fig:recrate_falc} we show $r_c$ for all snapshots of the NE-run.
The upper panel shows recombination from \HeII\ to \HeI. At the low temperature end ($T=5$~kK) there is a steep downturn and a horizontal extension toward higher temperatures at $r_2=0$. These are points in the photosphere. Between $T=20$~kK and $T=60$~kK there is net photoionization (negative $r_2$). We choose to model $r_2$ with its time average, except at the low temperature end, where we used an increase in recombination with decreasing temperature. We neglect the net photoionization between $T=20$~kK and $T=60$~kK, as there is very little neutral helium at those temperatures; the error we make is thus small. Similarly we ignore the downturn at the low temperature end. These points correspond to the photosphere where there is little \HeII\ and the high density there causes  the rates to be dominated by collisional processes.

The lower panel of Figure~\ref{fig:recrate_falc} shows $r_3$. There is very little spread, except in the low temperature end, which confirms the radiative recombination cascade scenario
illustrated in Figure \ref{fig:heii_transitions}. We chose the time-averaged values of $r_3$ for the 3-level model atom, with an extrapolation at the low temperature end.
 
\section{Results}\label{sec:results}
\subsection{Time-dependent  ionization with the full model atom}
\begin{figure*}
  \includegraphics[width=\textwidth]{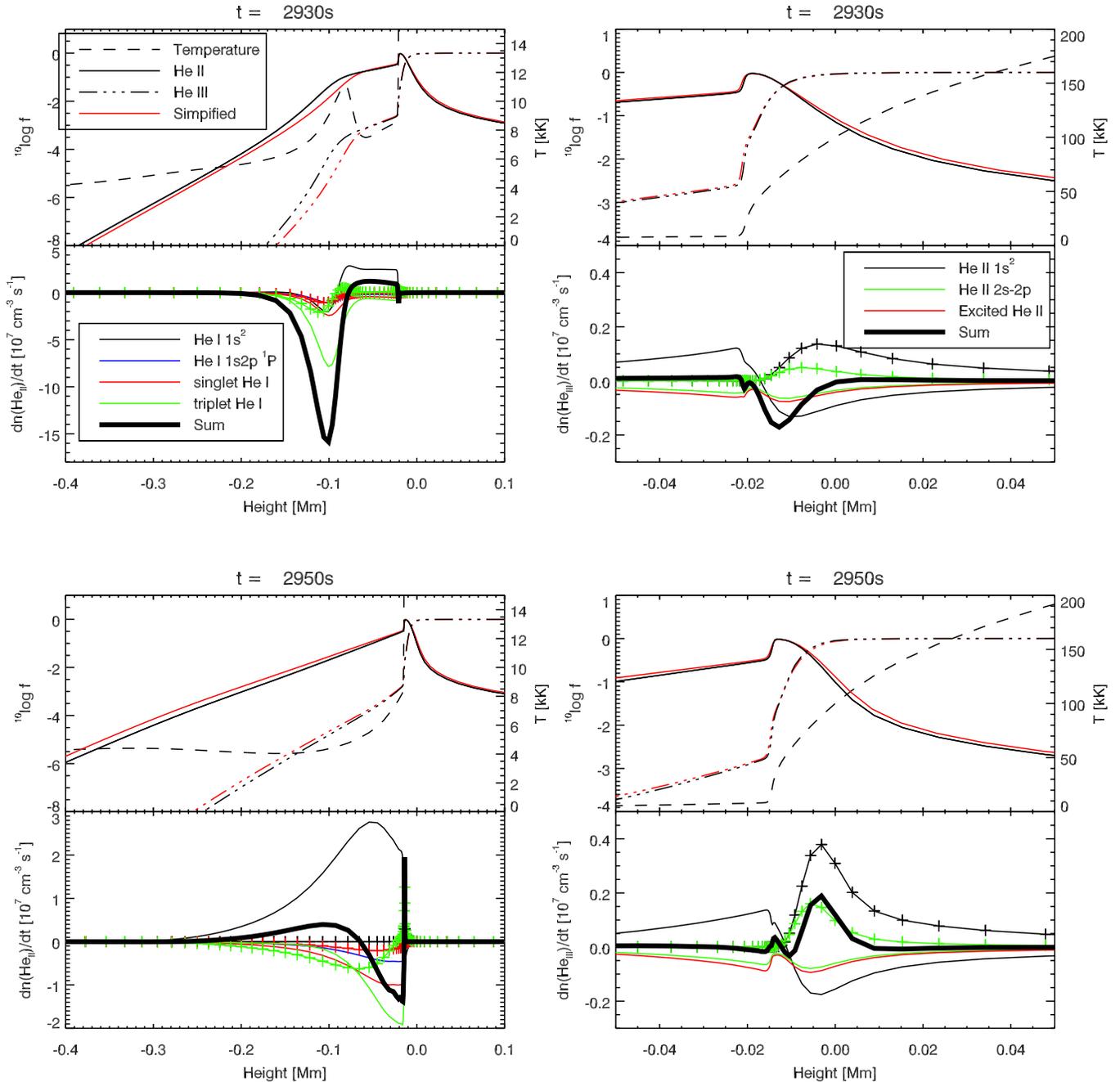}
 \caption{Ion fractions and net rates for two representative NE-run
  snapshots. Ion fractions from NE3-run are shown for comparison. 
  The lines with plus signs represent collisional processes.
  Each row corresponds to a snapshot. The left column shows
  the conditions in the chromosphere and rates into the ground state of
  \HeII. The right column shows the conditions in the transition 
  region and the rates into \HeIII. To be able to better compare the 
  structure of the transition region in the two runs, we have set
  $z=0$~Mm at $T=$100~kK.}
  \label{fig:processes}
\end{figure*}
As waves propagate upwards in the atmosphere, the thermodynamic conditions change
too fast for the ion densities to adjust, resulting in
non-equilibrium ion fractions.
Details of the ion fractions and net transition rates from two
snapshots of the NE-run are shown in Figure \ref{fig:processes}. The simulations
reveal, as far as the helium transitions are concerned, two types of
situations: pre-shock and post-shock, which are represented 
in the two chosen snapshots. 
In the chromospheric pre-shock phase ($t=2930$~s,
$z=[-0.08,-0.02]$~$\mathrm{Mm}$, where $z=0$~Mm is defined to be where the temperature is 100 kK) there is a net
ionization. This is driven by the EUV incident radiation which is
photoionizing neutral helium. Counteracting radiative recombination
to excited states is taking place, but the rates are not large
enough to balance the photoionization. 
When the shock passes and compresses the gas, the electron
density increases by both hydrogen ionizing and the compression
itself (CS2002). 
The radiative recombination rate coefficient
depends linearly on electron density, whereas the photoionization
rate coefficient is independent of electron density 
(Eqs. \ref{eq:photoionization}--\ref{eq:spontrec}). This
results in an increase of the net recombination in the post-shock
phase ($z<-0.08$~Mm). Collisions play only a minor role in the
chromosphere.
In the snapshot $t=2950$~s the post-shock phase is slowly adjusting
itself back to the pre-shock phase where photoionization dominates,
concurrent with the depletion of electrons 
which is due to the recombination of hydrogen.

The effects of \HeIII\ are most important in the transition region where
$T \sim 10-100$~kK, as the relative amount of neutral hydrogen and
helium here is small. Where $T<50$~kK, \HeII\ is behaving similarly to
\HeI: photoionizations from the ground state of \HeII,
driven by the EUV incident radiation, is roughly balanced by
radiative recombinations to the excited states. At $T>50$~kK, however,
collisional ionization from the ground state of \HeII\ replaces 
photoionization as the dominating rate, peaking at $T=90$~kK. Also 
note the non-negligible collisional ionization from the 
meta-stable \HeII\ $2s$ state.
In the transition region pre-shock phase
($t=2930$~s)  the radiative recombination dominates, causing a net 
recombination. When the shock passes, the
temperature rises due to the compression, and the higher temperature results
in net ionization as the net collisional ionization rates grow large ($t=2950$~s). 
When the plasma cools down, the net rates adjust themselves back towards 
the pre-shock phase.

\subsubsection{Time-dependent  ionization with the simplified model atom}
Ionization fractions obtained with the simplified model atom (from the
NE3-run) are shown together with the NE-run ion fractions in Figure
\ref{fig:processes}. We have replaced 30 atomic states in the
simplified model atom with two extra recombination
rates. Processes involving the removed states are not taken into
account, and this leads to a small shift in the position of the
transition region compared to the NE-run. We have set $z=0$~Mm where
$T=100$~kK in both simulations to be able to compare the
structure of the transition region directly from the figure.

The simplified model atom reproduces the ion fraction from the NE-run
very well, despite the neglected effects. Processes not taken into
account include 3-body recombination to- and
ionization from the excited states of both \HeI\ and \HeII. Figure
\ref{fig:processes} shows that these effects are subordinate.

\subsection{Relaxation timescales}\label{section:timescales}
\begin{figure}
  \includegraphics[width=\columnwidth]{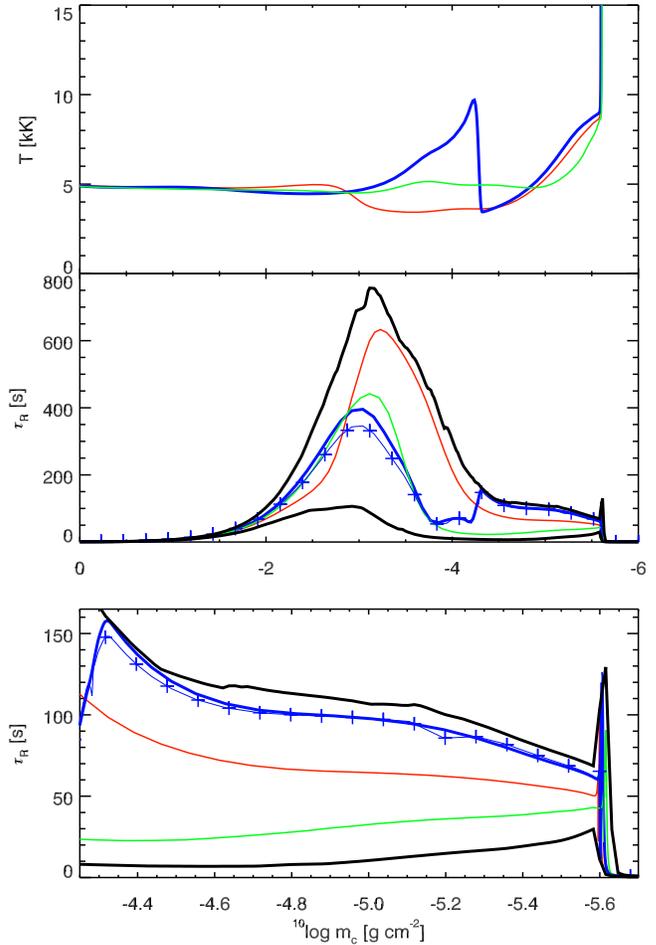}
  \caption{The temperature profiles of three snapshots (upper panel)
  and their corresponding relaxation timescales (middle
  and lower panel). The middle panel shows the whole computational domain, the lower panel is a blowup of the chromosphere and transition region.
  Each snapshot is shown in its own color (red, green and blue) in all panels.
  The maximum and minimum timescales obtained 
  are shown in thick black. The timescale obtained with the simplified
  model atom is show for one atmosphere only (thin blue curve with plus signs).}
  \label{fig:timescale}
\end{figure}

We define a relaxation timescale in the same way as in CS2002. As
described in Section \ref{sec:simulations}, we have initial ($n_0$),
time-dependent ($n(t)$) and final ($n_{\infty}$) population densities
from representative snapshots where the temperature is perturbed. The
relaxation timescales, $\tau_R$, are obtained numerically by fitting
the time-dependent population density of the ground state of \HeII\ to 
the general solution of a rate equation, assuming two levels
and constant transition rate coefficients:
\be
  n(t)=n_{\infty} - (n_{\infty} - n_{0})e^{-t/\tau_R}.
  \label{eq:timescales}
\ee 
Figure \ref{fig:timescale} shows the the relaxation timescales of the
three representative snapshots featuring both the pre-shock phase and
the post-shock phase. In the corona and photosphere, large rates result
in short relaxation timescales. The relaxation timescale is largest around the
photosphere/chromosphere interface at a column mass of
$^{10}\log{m_c}\approx -3$~g~cm$^{-2}$ with typically a few hundreds of
seconds. In the chromosphere it ranges from a couple of tens and up to
about a hundred seconds.

A wave front is building up at
$^{10}\log{m_c}=-2.7$~g~cm$^{-2}$ in the red atmosphere.  At this
stage, a shock has just passed through the transition region and the
chromospheric relaxation timescale is adjusting itself back to the
pre-shock values of about 60-100~s. The shock propagates through the
chromosphere (blue line) and reduces the timescale to its post-shock
value of a few tens of seconds (green line).

In the transition region at $^{10}\log{m_c}=-5.6$~g~cm$^{-2}$ there is
a sharp spike in $\tau_R$, marking the position of where the
ionization shifts from a balance between \HeI\ and \HeII\ to a balance
between \HeII\ and \HeIII. The definition of $\tau_R$ is based on a two
level system, and will not be very meaningful when three levels are
important at the same time. However, the spike itself is
within the region where the balancing is between \HeII\
and \HeIII, i.e. Eq. \ref{eq:timescales} is a decent approximation and
provides a meaningful estimate. 

The simplified model atom reproduces the relaxation timescales we 
get with the full 33-level atom very well. The model atom includes the
most important transition processes - clearly these are also the
dominant in setting the timescales of the system.

\subsubsection{Timescales of various processes}
We assume constant transition rates and use the analysis tools
developed in \cite{judge2005}. The set of rate equations in a
static atmosphere constitutes a set of first order differential
equations that can be expressed as a matrix equation:
\be
 \dot{{\bf n}} = P{\bf n},
 \label{eq:rateeq}
\ee  
with the solution
\be
  {\bf n} = \sum_i c_i {\bf v}_i e^{\lambda_i t},
  \label{eq:rateeq_solution}
\ee
where $P$ is the rate matrix, ${\bf n}$ is a vector containing the
population densities, ${\bf v}_i$ and $\lambda_i$ are the
corresponding eigenvectors and eigenvalues of the rate matrix and the
coefficients, $c_i$,
are constants depending on the initial conditions. The eigenvectors
represent different relaxation processes. One of the eigenvalues is 
zero and the corresponding eigenvector is proportional to the SE
solution. Each of the processes has a corresponding relaxation
timescale, given by $\tau_i=-1/\lambda_i$. The
smallest non-zero eigenvalue (in absolute value since they are all
negative) corresponds to the slowest relaxation process and thereby
determines the relaxation timescale of the system as a whole. 

By inspecting the eigenvectors in the 33-level model atom we find that
the balancing between the ground states of \HeI\ and \HeII\ 
is the slowest process in the chromosphere. The
second slowest, by a factor of 5-6, is the balancing between the
ground state of \HeII\ and \HeIII. This is
valid in both the pre and post-shock phases. In the transition region
where $T=60$~kK (corresponding to the peak in the
relaxation timescale, see Figure \ref{fig:timescale}) balancing between
the ground state of 
\HeII\ and \HeIII\ is the slowest
process. All other processes happen on timescales less
than a second.  

\subsection{Non-equilibrium effects}   
The relaxation timescales of helium in the chromosphere and transition
region ranges from a  few tens of seconds to about one hundred
seconds. When shock waves propagate through these regions the
hydrodynamic conditions may change too fast for the ionization balance
to adjust. Statistical equilibrium is an often-used assumption when
producing synthetic spectra, but this may not be a good assumption for
the spectral lines of helium if there are rapid changes
in the solar atmosphere. To obtain a feel for the types of
errors such an assumption would cause, we perform an initial
investigation of the non-equilibrium effects of \HeI\ 10830 and \HeII\ 304.

\subsubsection{\HeII\ 304}

\begin{figure*}
   \includegraphics[width=\textwidth]{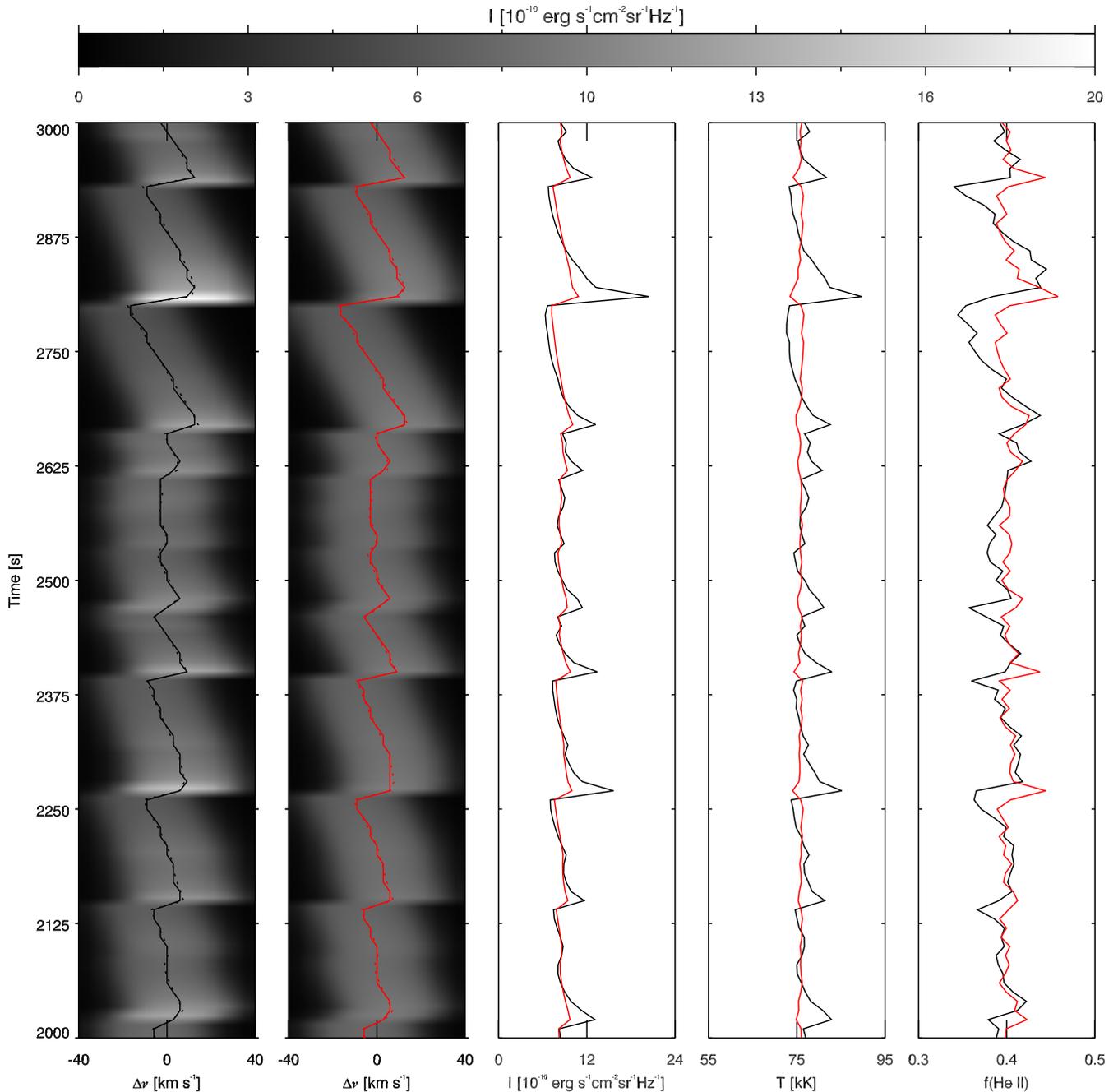}
  \caption{Formation of the \HeII\ 304
    line. From left to right: Emergent intensity in the NE-run, emergent intensity in the
    SE-run, line core intensities, temperature at  line core optical depth unity and fraction of
    \HeII\ at  line core optical depth unity ($\tau=1$). In the three rightmost
    panels the NE-run values are black and SE-run values are red.  The
    line core is defined as the wavelength where the emergent intensity is at
    its maximum and indicated by the solid lines in the two leftmost
    panels. Also indicated in these panels is  the plasma velocity at
    $\tau=1$ (dotted lines), which
    overlaps almost completely with the doppler shift of
    the line core.}
 \label{fig:304_int}
\end{figure*}

The resonance line \HeII\ 304 has two components, 
$1s\,^2\!S^\mathrm{e}_{1/2}-2p\,^{2}\!P^{\mathrm{o}}_{1/2,3/2}$  
(excluding the forbidden  
$1s\,^2\!S^\mathrm{e}_{1/2}-2s\,^{2}\!S^{\mathrm{e}}_{1/2}$), treated
separately in our simulations.
We investigated the
component with the largest oscillator strength ($1s
\,^2\!S^\mathrm{e}_{1/2}-2p\,^{2}\!P^{\mathrm{o}}_{3/2}$). 

Figure \ref{fig:304_int} compares the formation of the 304 line in the 
NE-run and the SE-run. A shockwave-induced sawtooth pattern is present in both runs, and
the intensity of the line generally increases as the shock front passes
the transition region where the line forms. 

We first discuss the NE line formation.
During the shock front passage the temperature increases on a
timescale shorter than the ionization-recombination timescale (see Section
\ref{section:timescales}), and the \HeII\ is therefore not ionized away. The compression work done by the shock is converted mainly into thermal energy and not to ionization energy, and the temperature increases strongly. 

Investigation of the rates show that the line
photons are mainly produced by collisional excitation from the ground
state. The large temperature rise leads to increased collisional excitation and thus strong emission in the 304 line as the shock passes the transition region. This is why the line core intensity and temperature at the
formation height are so strongly correlated.

In contrast, in the SE-run \HeII\  is instantaneously ionized as the shock front passes.
This has two
consequences, both of which reduce the number of 304 photons produced. First, the
temperature increase is smaller since the increase of energy goes into ionizing helium. Second, the instantaneous ionization to \HeIII\ lowers the 304 opacity, and shifts the
$\tau=1$ height down, where the temperature is lower. The line intensity in the SE-run is thus lower than in the NE-run. 
A good example of this can be seen at $t=2810$~s. 

The plasma velocity at the formation height nicely coincides with the
doppler shift of the line. This suggests that the line is formed in 
a thin atmospheric layer.

\subsubsection{\HeI\ 10830}

\begin{figure}
   \includegraphics[width=\columnwidth]{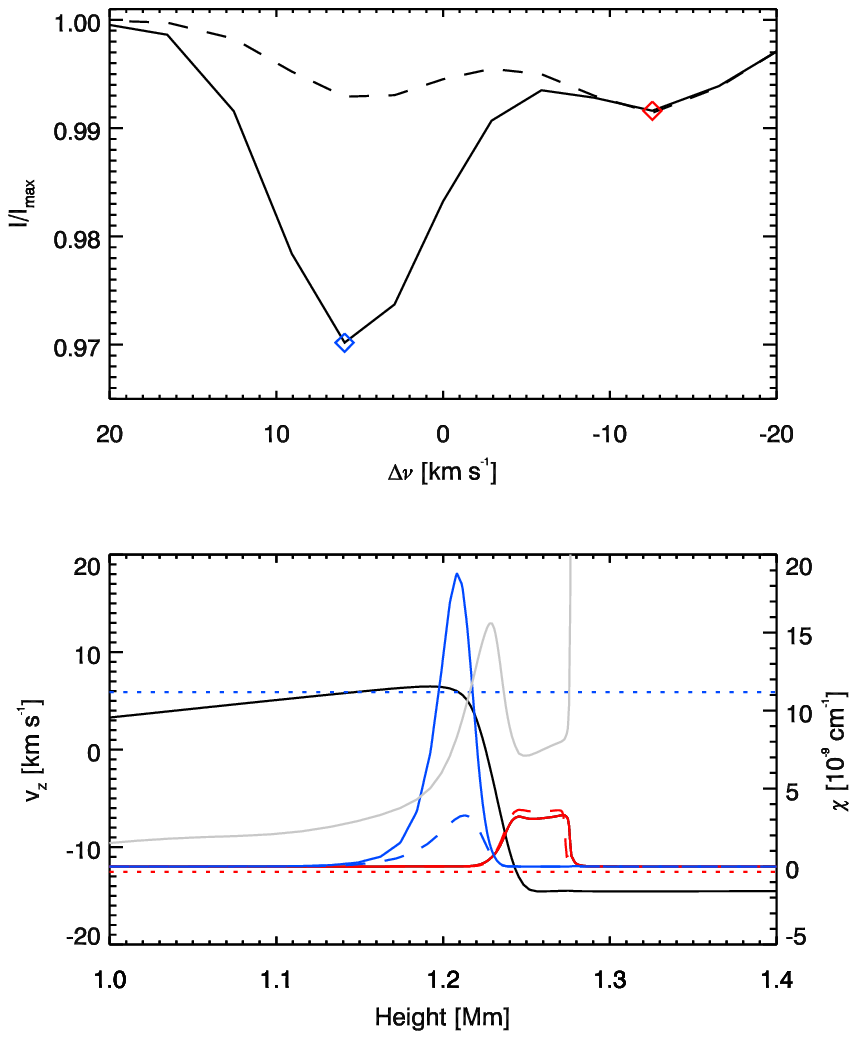}
  \caption{Formation of the \HeI\ 10830 line profile at
    $t=1650$~s. Upper panel: Emergent intensity for the NE-run (solid)
    and SE-run (dashed). The red and blue diamonds indicate the
    wavelengths used in the lower panel. Lower panel:  velocity in the
    atmosphere (solid black, scale to the left) and opacity at the two
    selected wavelengths (blue and red solid and dashed, scale to the
    right). The blue curves correspond to the frequency of the blue
    diamond in the upper panel (solid for NE, dashed for SE), the red
    curves for the frequency of the red diamond. The red and blue
    frequencies are also indicated on the velocity scale by the horizontal 
    dotted lines in the lower panel. In addition we show the temperature in grey.
    By coincidence the line profiles resemble the 10830
    blend, so we stress that the profiles shown here are only
    the strongest component of the line.
     The red
     dip in the line is formed in the pre-shock phase where matter is
     falling downward, and the blue dip is formed in the shock wave.}
 \label{fig:10830}
\end{figure}

The \HeI\ 10830 line forms when continuum photons from the
photosphere are scattered and absorbed in the chromosphere by neutral helium atoms 
residing in the $2s\,^{3}\!S^{\mathrm{e}}_1$ state in the triplet system
of \HeI. This is the lower level of the three 10830 transitions, $2s\,^{3}\!S^{\mathrm{e}}_1-
2p\,^{3}\!P^{\mathrm{o}}_{0,1,2}$.  

The triplet states are populated through recombination of \HeII\ 
\citep{avrett1994, mauas2005}. This
recombination happens on very short timescales compared to the
timescales of the ionization (see Section
\ref{section:timescales}). This means that the $2s\,^{3}\!S^\mathrm{e}_1$ 
population density, and hence the opacity, is always adjusted to the
amount of \HeII.
Our simulations treat each of the three components of
the 10830 line separately, and we choose to investigate the component
with the largest oscillator strength ($2s\,^{3}\!S^{\mathrm{e}}_1-
2p\,^{3}\!P^{\mathrm{o}}_{2}$). 

The upper panel of Figure \ref{fig:10830} shows the line profiles for both NE and SE runs at $t=1650$~s. The line profile has two depression cores at positive and negative doppler shift relative to the rest wavelength. The NE profile has a much deeper blue-shifted depression then the SE profile.

The lower panel shows the opacity at the wavelengths of the maximum depression together with the structure of the atmosphere. The red-shifted absorption is formed in the downflowing pre-shock material, the blue shifted component in the upflowing material behind the shock front. The doppler shift of the minima of the absorption components is nearly equal to the gas velocity in the line-forming region.

The opacity of the blue-shifted component in the SE-run is much lower
than in the NE-run, explaining the depth difference of the depression cores.
In this region
ionization equilibration timescales are around 50~s (see Figure
\ref{fig:timescale}). The helium population densities of the SE-run
equilibrates instantaneously to the increased electron density
associated with the shock, resulting in a smaller amount of
\HeII. Since \HeII\ is acting as a reservoir
for the triplet states, the 10830 opacity decreases
accordingly.

\section{Conclusions}\label{section:conclusions}
We have carried out several 1D radiation-hydrodynamic simulations in
order to study the processes determining helium ionization and the
timescales at which they work, and we have performed an initial investigation
of the formation of the \HeI\ 10830 and \HeII\ 304 lines.

Helium ionization is far from LTE in the upper chromosphere and 
transition region. Photoionization and
collisional ionization from the ground states are the determining 
processes in setting the level of helium ionization. These processes 
work on timescales of up to 100~s. Thermodynamic conditions are
sensitive to the ionization fraction of helium. Using an equation of
state with helium ionization assumed to be in LTE in numerical models will therefore give
erroneous results. To rectify this we have constructed a simplified
3-level helium model atom, based on the driving mechanisms, that 
reproduces the non-equilibrium ionization fractions quite well. The rate equations 
of the simplified model atom are simple enough for inclusion in 
multi dimensional stellar atmosphere codes. We plan to include
non-equilibrium ionization of helium in the 3D radiation-MHD code
Bifrost.  

We have shown that the 
formation of \HeI\ 10830 and \HeII\ 304 lines is sensitive to non-equilibrium effects.
Both lines show behavior that is not reproduced when SE is assumed. 
We therefore recommend exercising caution
when observations are interpreted  on the basis of SE computations.

\acknowledgements
This research was supported by the Research Council of Norway through the grant ``Solar Atmospheric
Modelling'' 
and by the European Research Council under the European Union's 
Seventh Framework Programme (FP7/2007-2013) / ERC Grant 
agreement nr. 291058. We would like to thank the referee for helpful comments
in the preparation of this manuscript.

\bibliographystyle{apj}
\bibliography{references}

\end{document}